\documentclass[aps,twocolumn,nofootinbib,superscriptaddress]{revtex4-1}
\usepackage{amsfonts,ulem,bbold}
\usepackage{amssymb,amsmath}
\usepackage{lipsum}
\usepackage{color}
\newcommand{\nn}{\nonumber}

\newcommand{\be}{\begin{equation}}  
\newcommand{\ee}{\end{equation}}  
\newcommand{\bea}{\begin{eqnarray}}  
\newcommand{\eea}{\end{eqnarray}}  

\newcommand{\R}{ \bar{R}} 
\newcommand{\e}{ \bar{e}}
\newcommand{\etil}{ \tilde{e}}

\newcommand{\N}{ \bar{N}} 
\newcommand{\Ntil}{ \tilde{N}} 
\newcommand{\vtil}{ \tilde{v}} 
\newcommand{\gtil}{ \tilde{\gamma}} 

\newcommand{\Comment}[1]{{}}
\definecolor{MyDarkBlue}{rgb}{0.15,0.15,0.45}
\usepackage[linktocpage=true]{hyperref}
\hypersetup{
colorlinks=true,
citecolor=MyDarkBlue,
linkcolor=MyDarkBlue,
urlcolor=MyDarkBlue,
}

\begin{document}

\title{A Note on Ghost-Free Matter Couplings in Massive Gravity and Multi-Gravity} 

\author{Kurt Hinterbichler}
\affiliation{Perimeter Institute for Theoretical Physics,\\
31 Caroline St. N, Waterloo, Ontario, Canada, N2L 2Y5}  
\author{Rachel A. Rosen}
\affiliation{Physics Department and Institute for Strings, Cosmology, and Astroparticle Physics,\\
Columbia University, New York, NY 10027, USA}

\begin{abstract}
We consider a recently proposed non-minimal matter coupling in massive gravity and multi-gravity.  We argue that, when formulated in terms of unconstrained vielbeins, this matter coupling contains the primary constraints necessary to remove the Boulware-Deser ghost to all orders away from the decoupling limit.
\end{abstract}  
  
\maketitle

\noindent
{\section{Introduction and Summary} }
\vspace{-.2cm}
Recent years have seen great progress in constructing and understanding effective field theories of massive gravity \cite{deRham:2010ik,deRham:2010kj,Hassan:2011hr} and its close cousins bi-gravity \cite{Hassan:2011zd} and multi-gravity \cite{Hinterbichler:2012cn} (see \cite{Hinterbichler:2011tt,deRham:2014zqa} for reviews).  Part of this quest has been to determine if, beyond the mass terms of de Rham, Gabadadze and Tolley (dRGT) \cite{deRham:2010kj}, there are other interactions in these theories that are free of the instability known as the Boulware-Deser ghost \cite{Boulware:1973my}.   In particular, there has been interest in non-minimal couplings to matter \cite{Hassan:2012wr,Akrami:2013ffa,deRham:2014naa,Noller:2014sta,Hassan:2014gta,deRham:2014fha,Soloviev:2014eea,Akrami:2014lja,Solomon:2014iwa,Gao:2014xaa,Yamashita:2014fga,Gumrukcuoglu:2014xba,Heisenberg:2014rka,Enander:2014xga,Schmidt-May:2014xla,Gumrukcuoglu:2015nua,Comelli:2015pua}.

In \cite{deRham:2014naa}, a coupling to matter using more than one metric was introduced.  It was argued that, although there is a phenomenologically interesting regime of validity for which the ghost is absent, this matter coupling causes the reappearance of the Boulware-Deser ghost above the strong coupling scale of the theory.  In this work we consider an analogous matter coupling expressed in terms of unconstrained vielbeins rather than metrics.    We argue that this version of the matter coupling preserves the primary constraint necessary to remove the Boulware-Deser ghost, to all orders beyond the decoupling limit.  In the presence of the matter coupling, the unconstrained vielbein formulation is not equivalent to the metric formulation, so there is no potential discrepancy between our results and those of \cite{deRham:2014naa}.

\vskip.4cm

\noindent
{\section{ Non-Canonical Matter Coupling}}
\vspace{-.2cm}

Consider the ghost-free bi-gravity theory of \cite{Hassan:2011zd} written in terms of two vielbeins $e_\mu^{~A}$ and $\e_\mu^{~A}$ as in \cite{Hinterbichler:2012cn}.  In $D$ dimensions, the theory includes an Einstein-Hilbert term for each vielbein,
\bea
&& \epsilon_{A_1\cdots A_D} \, R^{A_1A_2}\wedge e^{A_3} \wedge\cdots\wedge e^{A_D}\, ,  \nn \\
&& \epsilon_{A_1\cdots A_D} \, \R^{A_1A_2}\wedge \e^{A_3} \wedge\cdots\wedge \e^{A_D}\, ,  \nn 
\eea
where $R^{AB}$ and $\R^{AB}$ are the curvature two-forms corresponding to $e_\mu^{~A}$ and $\e_\mu^{~A}$ respectively.  In addition, the ghost-free bi-gravity theory contains a linear combination of $D+1$ non-derivative terms \cite{deRham:2010kj,Hassan:2011hr,Hassan:2011zd,Hinterbichler:2012cn},
\be 
\epsilon_{A_1\cdots A_D} \, e^{A_1} \wedge\cdots\wedge e^{A_n} \wedge \e^{A_{n+1}} \wedge \cdots \wedge \e^{A_D},
\ee 
for $n=0,1,\cdots, D$.  This first and last of these are cosmological constants, while the others induce genuine interactions among the vielbeins.

Let us couple this theory to some scalar matter sector that does not contain higher derivative terms.  In particular, let us consider minimal coupling to a composite vielbein $\etil_\mu^{~A}$ that is a function of the two vielbeins $e_\mu^{~A}$ and $\e_\mu^{~A}$.  Diffeomorphism invariance of the matter action ensures that the matter Hamiltonian is linear in the effective lapse and shift variables,
\be
{\cal H}_{matter} = \Ntil {\cal C} + \Ntil^i {\cal C}_i \, ,
\ee
with some $ {\cal C}, {\cal C}_i$ depending on the matter fields, their canonical momenta and the spatial metric variables.

If the Hamiltonian is also linear in the original lapse and shift variables $N$, $N^i$, $\N$ and $\N^i$, then we can apply the arguments of \cite{Hinterbichler:2012cn}, and conclude that this matter coupling, when taken together with the bi-gravity theory in terms of $e_\mu^{~A}$ and $\e_\mu^{~A}$, will preserve the primary constraint necessary to remove the Boulware-Deser ghost.  

Here we argue that this is the case if the effective vielbein is a linear combination of the two other vielbeins,
\be
\etil_\mu^{~A} = e_\mu^{~A}+ \alpha\, \e_\mu^{~A} \, .\label{lincomb}
\ee
Here $\alpha$ is a free constant parameter and we have, without loss of generality, set the coefficient of $e_\mu^{~A}$ in the linear combination to unity by a re-scaling.  Such a coupling was introduced in \cite{deRham:2014naa} and studied in the metric language.    The equivalent formulation in terms of vielbeins was introduced in \cite{Noller:2014sta}.

For the unconstrained vielbein, we adopt a boosted ADM \cite{Arnowitt:1960es,Peldan:1993hi} decomposition, following \cite{Hinterbichler:2012cn}:
\be
e_\mu^{\ A} =
\left(\begin{array}{cc}N \gamma +N^i e_i^{\ a} v_a& N v^a+N^i e_i^{\ b} (\delta_b^{\ a}+\frac{1}{\gamma+1}v_b v^a) \\
e_i^{\ a} v_a &  e_i^{\ b} (\delta_b^{\ a}+\frac{1}{\gamma+1}v_b v^a)\end{array}\right) \, , 
\ee 
\be
\e_\mu^{\ A} =
\left(\begin{array}{cc}\N & \N^i \e_i^{\ a} \\
0  &  \e_i^{\ a} \end{array}\right) \, .
\ee 
Here, $N$, $N^i$, $\N$ and $\N^i$ are the usual lapse and shift for $e_\mu^{~A}$ and $\e_\mu^{~A}$ respectively.  The variables $e_i^{\ a}$ and $\e_i^{\ a}$ are the degrees of freedom associated with the spatial component of the vielbeins.  $v^a$ represents the parameter associated with a Lorentz boost and $\gamma \equiv \sqrt{1+v^a v_a}$.  We have used the overall Lorentz invariance of the bi-gravity theory to set ${\bar v}^a =0 $.  
For the composite vielbein, we write
\be
\etil_\mu^{\ A} =
\left(\begin{array}{cc} \Ntil \gtil +\Ntil^i \etil_i^{\ a} \vtil_a& \Ntil \vtil^a+\Ntil^i \etil_i^{\ b} \left(\delta_b^{\ a}+\frac{1}{\gtil+1}\vtil_b \vtil^a\right) \\
\etil_i^{\ a} \vtil_a & \etil_i^{\ b} \left(\delta_b^{\ a}+\frac{1}{\gtil+1}\vtil_b \vtil^a\right)\end{array}\right) \, .
\ee

From \eqref{lincomb}, we can read off four equations that determine $\Ntil$, $\Ntil^i$, $\etil_i^{\ a}$ and $\vtil^a$:
\bea
&& \etil_i^{\ a} \vtil_a = e_i^{\ a} v_a  \, , \nn\\
&& \etil_i^{\ b} \left(\delta_b^{\ a}+\frac{1}{\gtil+1}\vtil_b \vtil^a\right) = e_i^{\ b} \left(\delta_b^{\ a}+\frac{1}{\gamma+1}v_b v^a\right)+ \alpha \, \e_i^{\ a} \, ,\nn \\
&& \Ntil \gtil +\Ntil^i \etil_i^{\ a} \vtil_a = N \gamma +N^i e_i^{\ a} v_a+ \alpha\, \N \, ,\nn \\
&& \Ntil \vtil^a+\Ntil^i \etil_i^{\ b} \left(\delta_b^{\ a}+\frac{1}{\gtil+1}\vtil_b \vtil^a\right) = \nn \\
&&~~~~~~~~~~~ N v^a+N^i e_i^{\ b} \left(\delta_b^{\ a}+\frac{1}{\gamma+1}v_b v^a\right) + \alpha\,  \N^i \e_i^{\ a}\, . \label{NNi}
\eea
The key point is that the first two equations of \eqref{NNi} are entirely independent of the lapse and shift variables, and the second two are linear in the lapse and shift variables.  The first two can be solved for $\etil_i^{\ a}$ and $\vtil^a$ and the result will depend only on $e_i^{\ a}$, $\e_i^{\ a}$ and $v^a$.   Plugging the solution into the second two and solving for $\Ntil$ and $\Ntil^i$, the solutions for $\Ntil$ and $\Ntil^i$ will be linear in $N$, $N^i$, $\N$ and $\N^i$, as desired.   

Following \cite{Hinterbichler:2012cn}, we can use the $\N^i$ equation of motion to eliminate the boost parameter $v^a$ and get an action that remains linear in $N$, $N^i$ and $\N$.  $N$, $N^i$ then enforce the first class primary constraints of overall diffeomorphism invariance, and $\N$ enforces the extra primary constraint that eliminates the ghost.

In $D=2$ we can be more explicit.  The zweibeins are parametrized as
\bea
e_\mu^{~A} &=& \left(\begin{array}{cc}N\sqrt{1+v^2} +N^1\, e \,v & N\, v+N^1\,e\,\sqrt{1+v^2} \\e\,v & e\, \sqrt{1+v^2}\end{array}\right) \, , \nn\\
\e_\mu^{~A} &=& \left(\begin{array}{cc}\N  & \N^1\,\e \\0 & \e \end{array}\right) \, , \nn\\
\tilde e_\mu^{~A} &=& \left(\begin{array}{cc}\tilde N\sqrt{1+\tilde v^2} +\tilde N^1\, \tilde e \,\tilde v & \tilde N\, \tilde v+\tilde N^1\,\tilde e\,\sqrt{1+\tilde v^2} \\ \tilde e\,\tilde v & \tilde e\, \sqrt{1+\tilde v^2}\end{array}\right) \, .\nn\\
\eea
Equations \eqref{NNi} become
\bea
&&\etil\, \vtil =  e\, v  \, , \nn\\
&&\etil \, \sqrt{1+\vtil^2}  = e \, \sqrt{1+v^2}  +\alpha \, \e\, , \nn\\
&& \Ntil\sqrt{1+\vtil^2} +\Ntil^1\, \etil \,\vtil   = N\sqrt{1+v^2} +N^1\, e \,v+ \alpha \N \, , \nn\\
&& \Ntil\, \vtil+\Ntil^1\,\etil\,\sqrt{1+\vtil^2}  = N\, v+N^1\,e\,\sqrt{1+v^2}+\alpha \N^1 \e \, . \nn\\
 \label{2desprss}
\eea
Solving the first two of \eqref{2desprss}, $e$ and $v$ are independent of all lapses and shifts:
\bea
\etil &=& \sqrt{e^2+\alpha^2\e^2+2\alpha\,e\,\e\,\sqrt{1+v^2}} \, , \nn \\
\vtil &=& \frac{e\, v}{\sqrt{e^2+\alpha^2\e^2+2\alpha\,e\,\e\,\sqrt{1+v^2}}} \, .
\eea
Using this in the final two expressions of \eqref{2desprss}, we have
\be \footnotesize
\Ntil = \frac{N \,e +\alpha^2\,\N\,\e+\alpha\,(N^1-\N^1)v\,e\,\e+\alpha\,(N\,\e+\N\,e)\,\sqrt{1+v^2}}
{\sqrt{e^2+\alpha^2\e^2+2\alpha\,e\,\e\,\sqrt{1+v^2}}} \, ,
\ee
\be \footnotesize
\Ntil^1 = \frac{N^1 \,e^2 +\alpha^2\N^1\,\e^2+\alpha\,(N\,\e-\N\,e)v+\alpha\,(N^1+\N^1)\,e\,\e\sqrt{1+v^2}}{e^2+\alpha^2\e^2+2\alpha\,e\,\e\,\sqrt{1+v^2}} \, .
\ee
Thus the lapse and shift of the composite zweibein remain linear in the lapses and shifts of the two original vielbeins.  Indeed, in $D=2$ dimensions the argument for ghost-freedom is ultimately trivial since the non-canonical mater coupling can always be field-redefined away without affecting the structure of the Einstein-Hilbert kinetic terms or the ghost-free mass terms.  

For dimensions greater than two, the system of equations \eqref{NNi} can be solved when the matrix $M_i^{\\ a} \equiv e_i^{\ b} \left(\delta_b^{\ a}+\frac{1}{\gamma+1}v_b v^a\right)+ \alpha \, \e_i^{\ a}$ is invertible.  Then, the linearity in the lapse and shift variables will guarantee a primary constraint that is necessary to eliminate the Boulware-Deser ghost to all orders.  The existence of an associated secondary constraint must also be demonstrated for these theories to be consistent at all scales.

\vskip.4cm

\noindent
{\section{Vielbeins vs Metrics} }  
\vspace{-.2cm}
Our conclusion is not inconsistent with the claims of \cite{deRham:2014naa}, in which the metric formulation was assumed.  This is because the non-canonical matter coupling breaks the usual equivalence between the vielbein and metric formulations that allows one to make the replacement $e^{-1} \e \rightarrow \sqrt{g^{-1}\bar{g}}$.  In other words, massive gravity theories and bigravity theories with Einstein-Hilbert kinetic terms expressed in terms of vielbeins possess an on-shell constraint of the form
\be
\label{Uconst}
\frac{\partial U}{\partial e_\mu ^{\ A}} e_\mu^{\ B} \eta_{BC} = \frac{\partial U}{\partial e_\mu ^{\ C}} e_\mu^{\ B} \eta_{BA} \, ,
\ee
where $U$ is the part of the Lagrangian that contains no derivatives on the the vielbein (see, e.g., \cite{Hinterbichler:2012cn}).  This constraint is due to the overall Lorentz invariance of the kinetic terms and is responsible (in $D=4$) for removing the six spurious components of the vielbein.  For dRGT massive gravity theories and for the related bigravity theories with canonical matter couplings, this constraint equation can be solved by imposing the following condition on the vielbein:
\be
\label{Uconstsol}
e_{[\mu}^{\ A} \e_{\nu]}^{\ B} \eta_{AB} = 0 \, .
\ee
This condition can then be used to show that 
\be
e^\mu_{\ A} \e_\lambda^{\ A} e^\lambda_{\ B} \e_\nu^{\ B}  = g^{\mu\lambda}\bar{g}_{\lambda \nu} \, ,
\ee
where $g_{\mu\nu} = e_\mu^{\ A} e_\nu^{\ B} \eta_{AB}$ and $\bar{g}_{\mu\nu} = \e_\mu^{\ A} \e_\nu^{\ B} \eta_{AB}$.  Thus one can pass easily between the vielbein and metric formulations by making the replacement
\be
e^\mu_{\ A} \e_\nu^{\ A} \rightarrow \sqrt{g^{-1}\bar{g}}^\mu_{\ \nu} \, .
\ee
In the presence of the non-canonical matter coupling, one still has the constraint equation \eqref{Uconst}. However, \eqref{Uconstsol} will no longer be a solution.  This is because the non-derivative part of the Lagrangian $U$ will now contain matter fields.  Thus the solution to the constraint is dynamically modified in the presence of the double matter coupling, and therefore the equivalent metric formulation will be quite complicated and different from that of \cite{deRham:2014naa}. 

Said another way, to pass from the vielbein to the metric formulation, in the case with no matter coupling or with canonical matter coupling, we can parametrize the vielbein as a Lorentz boost times a symmetric vielbein \eqref{Uconstsol} .  The Lorentz boost parameters then appear without derivatives, and in such a way that they are set to zero by their own equations of motion (see \cite{Hinterbichler:2012cn,Deffayet:2012zc} for more details).  In particular, the Lorentz boost parameters do not appear in the canonical matter coupling, since it is locally Lorentz invariant under each of the two local Lorentz transformations.  In the case of doubly coupled matter, the matter coupling is no longer invariant under both local Lorentz rotations -- only the overall one -- so the Lorentz boost parameters appear in the matter sector, and the solution to their equations of motion can involve the matter fields.  

Thus, the unconstrained vielbein formulation and the metric formulation of \cite{deRham:2014naa} are different theories. 
We argue that only the unconstrained vielbein theory, in which the symmetrization constraint is determined dynamically, contains the primary constraints necessary for the theory to be ghost-free. Choosing to impose the usual symmetrization constraint, as in e.g. \cite{Hassan:2014gta,Soloviev:2014eea}, rather than letting it be determined dynamically in the presence of matter, results in a theory which is equivalent to the ghostly metric formulation of \cite{deRham:2014naa}.
Note, however, that because the ghost in the metric formulation appears above the strong coupling scale of the effective theory, this formulation is also acceptable
if one takes the strong-coupling scale to be the cutoff at which new physics enters.  It is only if
we allow for the possibility that the strong coupling scale is not a new-physics cutoff and that the
vielbein description has a strongly-coupled range of validity above the mass
of the would-be ghost of the metric formulation that the difference plays
a role\footnote{We are grateful to Claudia de Rham for discussions on this.}.

We note also that the constraint equation \eqref{Uconst} may have more than one branch of solutions.  Which branch of solutions one picks can affect whether or not there are valid secondary constraints that fully remove the Boulware-Deser ghost.  In other words, while all theories of this form contain the primary constraints, not all branches of solutions to \eqref{Uconst} will necessarily be ghost-free.

\vskip.4cm

\noindent
{\section{Generalizations} }  
\vspace{-.2cm}
We have focused on the bi-gravity case, but our arguments go through in massive gravity as well, for which the second metric is frozen to some fiducial metric.   In addition, it is straightforward to extend our argument to matter couplings in theories of many interacting spin-2 fields $e_{\mu (I)}^{ \ A}$.  Such generalizations were also considered in \cite{Noller:2014sta}.  The effective vielbein is given by the linear combination
\be
\etil_\mu^A = \sum_I \alpha_{(I)} e_{\mu (I)}^{ \ A} \, .
\ee
The lapse and shift of the composite vielbein defined in this way will be linear in the lapses and shifts of the constituent vielbeins.

\vskip.4cm

\noindent
{\section{Outlook} }  
\vspace{-.2cm}
We have seen that, in the unconstrained vielbein formulation, bi-gravity and multi-gravity theories with matter multiply coupled to a linear combination of the various vielbeins have the primary constraint which is necessary to remove the Boulware-Deser ghost to all orders.  Because these couplings are not dynamically equivalent to the analogous couplings in the metric formulation, their cosmological and solar system phenomenology may be different and interesting to study.

This work demonstrates only the existence of primary constraints.  In order for the matter couplings considered here to be truly ghost-free to all orders, it must be demonstrated that they contain the appropriate secondary constraints as well.  Recent work that has appeared since the initial posting of this paper suggests that, in fact, this is not the case \cite{deRham:2015cha,Matas:2015qxa}.  Nevertheless, it has been emphasized that, while these new couplings may not be ghost-free at all energy scales, they remain viable below the cut-off of an effective theory.

\vskip.3cm

\noindent
{\bf Acknowledgements:} We are grateful to Claudia de Rham, Gregory Gabadadze, Garrett Goon, Lavinia Heisenberg, Austin Joyce, Johannes Noller, Riccardo Penco, Raquel Ribeiro, Andrew Tolley and Mark Trodden for many productive conversations.  Research at Perimeter Institute is supported by the Government of Canada through Industry Canada and by the Province of Ontario through the Ministry of Economic Development and Innovation. This work was made possible in part through the support of a grant from the John Templeton Foundation. The opinions expressed in this publication are those of the author and do not necessarily reflect the views of the John Templeton Foundation (KH).  RAR is supported by DOE grant DE-SC0011941.

\bibliographystyle{apsrev4-1}
\bibliography{matter_arxiv_v2}
\end{document}